\documentclass[12pt]{article}
\usepackage{amssymb,amsmath,epsfig}

\begin{document}

\title{\bf Non-Commutative Correction to Thin Shell Collapse in Reissner Nordstr$\ddot{o}$m Geometry}

\author{Muhammad SHARIF \thanks{msharif.math@pu.edu.pk} and Ghulam ABBAS
\thanks{abbasg91@yahoo.com}\\
Department of Mathematics, University of the Punjab,\\
Quaid-e-Azam Campus, Lahore-54590, Pakistan.}

\date{}
\maketitle

\begin{abstract}
This paper investigates the polytropic matter shell collapse in the
non-commutative Reissner-Nordstr$\ddot{o}$m geometry. Using the
Israel criteria, equation of motion for the polytropic matter shell
is derived. In order to explore the physical aspects of this
equation, the most general equation of state,
$p=k{\rho}^{({1+\frac{1}{n}})}$, has been used for finite and
infinite values of $n$. The effective potentials corresponding to
the equation of motion have been used to explain different states of
the matter shell collapse. The numerical solution of the equation of
motion predicts collapse as well as expansion depending on the
choice of initial data. Further, in order to include the
non-commutative correction, we modify the matter components and
re-formulate the equation of motion as well as the corresponding
effective potentials by including non-commutative factor and charge
parameter. It is concluded that charge reduces the velocity of the
expanding or collapsing matter shell but does not bring the shell to
static position. While the non-commutative factor with generic
matter favors the formation of black hole.
\end{abstract}

\section{Introduction}

Over the past decades, an extensive development has been made on the
Penrose Cosmic Censorship Hypothesis (CCH).$^{{1})}$ According to
this hypothesis, the singularities arising from the gravitational
collapse of regular initial data are always invisible to the far
away observer as these are clothed by the event horizons of black
hole. On the other hand, if a naked singularity is formed as end
state of the collapse, then it is visible to the external observer.
Since black hole and naked singularity have different properties, it
becomes important to explore the problem of collapse in order to
have a satisfactory asnwer of the CCH. Depending on the choice of
initial data, the gravitational collapse of a realistic matter leads
to the formation of black hole or naked singularity.$^{{2})}$
Virbhadra and his collaborators $^{{3}-{{8})}}$ introduced the idea
of gravitational lensing to determine the nature of singularity. In
one of his papers$^{{9})}$, he presented an improved form of the
CCH, which is a source of inspiration for researchers.

The study of gravitational collapse requires that one must consider
the appropriate geometry of the interior and exterior regions of a
star and junction conditions which allow the smooth matching of
these regions. According to Israel$^{{10})}$, there are two types of
the boundary surface one with thin layer of surface matter
distribution and the other without such thin layer. The matching
conditions over the second type of the boundary surface demands the
continuity of the metric coefficients as well as the extrinsic
curvature components. This leads to the classical model of the
relativistic gravitational collapse formulated by Oppenheimer and
Snyder.$^{{11})}$ On the other hand, matching the interior and
exterior regions over a boundary surface with thin layer of matter
requires the discontinuity of the extrinsic curvature over the
boundary surface. This approach has been used successfully for more
than thirty years to analyze the nature of singularity in the
relativistic gravitational collapse.

Pereira and Wang$^{{12})}$ studied the gravitational collapse of
cylindrical shells made of counter rotating dust particles by using
the Israel thin shell formalism. Sharif and Ahmad$^{{13})}$ have
extended this work to plane symmetric spacetime. Recently, Sharif
and his collaborators$^{{14}),{15})}$ have investigated the
spherically symmetric gravitational collapse for a class of
spacetimes using Israel thin shell formalism. This approach was
generalized to thin charged shell without pressure by de Ia Cruz and
Israel.$^{{16})}$ Kuchar$^{{17})}$ and Chase$^{{18})}$ are among the
first authors who treated the charged thin shell problem with
pressure by using the polytropic equation of state. There are a
number of papers devoted to handle the charged thin shell problems.
Boulware$^{{19})}$ studied the time evolution of the charged thin
shell and showed that their collapse can form a naked singularity.

According to Doplicer et al.$^{{20})}$, when the energy density
becomes sufficiently large, a black hole is formed. Heisenberg
uncertainty principle states that measurement of a spacetime
separation causes an uncertainty in momentum, i.e., momentum is
inversely proportional to the extent of the separation. When the
separation is small enough (as in the Schwarzschild spacetime,
$r\rightarrow 2m$), the system leads to the formation of black hole
which prevents any information escaping from the system. Thus there
is a lower bound for the measurement of length. The condition for
preventing the gravitational collapse can be expressed as
uncertainty relation for the coordinates. This relation can be
derived from the commutation relation of the coordinates.

There has been a growing interest to study the behavior of black
hole in the non-commutative field theory. Nicolini et al.$^{{21})}$
examined the behavior of the non-commutative Schwarzschild black
hole while Modesto and Nicolini$^{{22})}$ discussed the charged
rotating non-commutative black hole. Bastos et al. $^{{23}),{24})}$
explored the non-canonical phase space, singularity problem and
black hole in the context of non-commutativity. Recently, Bartolami
and Zarro$^{{25})}$ investigated the non-commutative correction to
pressure, particle numbers and energy density for fermion gas and
radiations. The non-commutative correction to these quantities lead
to the fact that non-commutativity affects the matter dispersion
relation and equation of state. Inspired by the non-commutative
correction to black hole physics, Oh and Park$^{{26})}$ explored the
gravitational collapse of shell with smeared gravitational source in
the non-commutative Schwarzschild geometry.

This paper extends the above work to the non-commutative RN
geometry. The main purpose is to investigate the effects of the
charge parameter and the non-commutative factor on the gravitational
collapse. The plan of the paper is as follows: In the next section,
we discuss the shell collapse for the RN spacetime. The
non-commutative correction to the shell collapse is presented in
section \textbf{3}. In the last section, we summarize and conclude
the results.

\section{Thin Shell Collapse}

We assume that the $3D$ timelike boundary surface ${\Sigma}$ splits
the two $4D$ spherically symmetric spacetimes $M^+$ and $M^-$. The
internal $M^-$ and external $M^+$ regions are described by the RN
metrics given by
\begin{equation}\label{1}
(ds)_\pm^2=L_{\pm} dT^2-\frac{1}{L}_\pm
dR^2-R^2(d\theta^2+\sin^2{\theta}d\phi^2),
\end{equation}
where $L_{\pm}(R)=1-\frac{2M_\pm}{R}+\frac{Q_\pm^2}{R^2},~M_\pm$ and
$Q_\pm$ are the mass and charge, respectively. The subscripts $+$
and $-$ represent the quantities in exterior and interior regions
respectively to the boundary surface ${\Sigma}$. Further, we assume
that charge in both the regions is same i.e., $Q_-= Q_+ =Q $. The
strength of the electric field on the shell can be described by the
Maxwell field tensor $F_{TR}=\frac{Q}{R^2}=-F^{RT}$. The
corresponding energy-momentum tensor of the electromagnetic field is
\begin{equation}\label{2}
{T_{\mu}^{\nu}}^{(em)}=\frac{1}{4{\pi}}
(-F^{{\nu}{\lambda}}F_{{\mu}{\lambda}}+\frac{1}{4}\delta^{\nu}_{\mu}
F_{{\pi}{\lambda}}F^{{\pi}{\lambda}}).
\end{equation}

By employing the intrinsic coordinates $(t,\theta, \phi)$ on the
${\Sigma}$ at $R=R(t)$, the metrics in Eq.(\ref{1}) become
\begin{equation}\label{3}
{(ds)_\pm^2}_\Sigma=[L_{\pm}(R)(\frac{dT}{dt})^2-\frac{1}{L_{\pm}(R)}
(\frac{dR}{dt})^2]dt^2-R(t)^2(d\theta^2+\sin^2{\theta}d\phi^2),
\end{equation}
here we assume that $g_{00}>0$, so that $T$ is a timelike
coordinate. Also, the induced metric on the ${\Sigma}$ is
\begin{equation}\label{4}
{(ds)^2}_\Sigma=dt^2-a(t)^2(d\theta^2+\sin^2{\theta}d\phi^2).
\end{equation}
The continuity of the first fundamental form gives
\begin{eqnarray}\label{5}
[L_{\pm}(R_\Sigma)-\frac{1}{L_{\pm}(R_\Sigma)}
(\frac{dR_\Sigma}{dT})^2]^{\frac{1}{2}}dT=(dt)_\Sigma,\\\label{c7}R(t)=a(t)_{\Sigma}.
\end{eqnarray}
The unit normal ${n_\mu}^\pm$ to the ${\Sigma}$ in $M^{\pm}$
coordinates can be evaluated as
\begin{eqnarray}\label{6}
{n_\mu}^\pm=(-\dot{R}(t),\dot{T}, 0, 0),
\end{eqnarray}
where dot represents differentiation with respect to $t$.

The extrinsic curvature tensor $K^\pm_{ij}$ on the $\Sigma$ is
defined as
\begin{eqnarray}\label{7}
K^\pm_{ij}={n_\sigma}^\pm(\frac{\partial^2x^{\sigma}_\pm}{\partial\xi^i\partial\xi^j}
+\Gamma^{\sigma}_{\mu\nu}\frac{\partial
{x_\pm}^\mu}{\partial\xi^i}\frac{\partial
{x_\pm}^\nu}{\partial\xi^j}),\quad (i,j=0,2, 3).
\end{eqnarray}
Using this expression, one can find the following non-vanishing
components of the extrinsic curvature
\begin{eqnarray}\label{8}
K^\pm_{tt}=\frac{d}{dR}\sqrt{\dot{R}^2+L_\pm},\quad
K^\pm_{\theta\theta}=-R\sqrt{\dot{R}^2+L_\pm},\quad
K^\pm_{\phi\phi}=K^\pm_{\theta\theta}\sin^2{\theta}.
\end{eqnarray}
The surface energy-momentum tensor is defined by
\begin{equation}\label{9}
S_{ij}=\frac{1}{\kappa}\{[K_{ij}]-\gamma_{ij}[K]\} ,
\end{equation}
where ${\kappa}$ is the coupling constant, $\gamma_{ij}$ is the
induced metric on the $\Sigma$ and
\begin{equation}\label{10}
[K_{ij}]=K^+_{ij}-K^-_{ij},\quad\quad [K]=\gamma^{ij}[K_{ij}].
\end{equation}

The surface energy-momentum tensor for a fluid of density $\rho$ and
pressure $p$ is
\begin{equation}\label{11}
S_{ij}=\rho\omega_i\omega_j+ p(\theta_i\theta_j+\phi_i\phi_j),
\quad(i,j=t, \theta, \phi),
\end{equation}
where $\omega_i,~\theta_i$ and $\phi_i$ are unit vectors defined on
the $\Sigma$ by the relations
\begin{equation}\label{12}
\omega_i=\delta^t_i,\quad \theta_i=a(t)\delta^\theta_i,
\quad\phi_i=a(t)\delta^\phi_i\sin\theta.
\end{equation}
Using Eqs.(\ref{c7}), (\ref{9}), (\ref{11}) and (\ref{12}), we can
find
\begin{equation}\label{13}
\rho=\frac{2}{\kappa R^2}[K_{\theta\theta}],\quad
p=\frac{1}{\kappa}\{[K_{tt}]-\frac{[K_{\theta\theta}]}{R^2}\}.
\end{equation}
With the help of the non-zero extrinsic curvature components, we get
\begin{eqnarray}\label{15}
(\zeta_+-\zeta_{-})+\frac{\kappa}{2}\rho R=0,
\\\label{14}
\frac{d}{dR}(\zeta_+-\zeta_{-})+\frac{1}{R}(\zeta_+-\zeta_{-})-{\kappa}p=0,
\end{eqnarray}
where $\zeta_\pm=\sqrt{\dot{R}^2+L_\pm}$. Equations (\ref{15}) and
(\ref{14}) can be reduced to the following single equation
\begin{equation}\label{16}
\frac{d\rho}{d\log R}+2(\rho+p)=0.
\end{equation}

The equation of state for the polytropic matter is
\begin{equation}\label{17}
p=k{\rho}^{({1+\frac{1}{n}})},
\end{equation}
where $k$ is the equation of state parameter and $n$ denotes the
polytropic index. Notice that different values of $n$ correspond to
different types of matter, for example, for $n\rightarrow\infty$, we
have perfect fluid. The solution of Eq.(\ref{16}), by using
(\ref{17}), for finite and infinite values of $n$  are
\begin{eqnarray}\label{18}
\rho=\{(k+{\rho_0}^{\frac{-1}{n}})(\frac{R}{R_0})^{\frac{2}{n}}-k\}^{-n},
\\\label{14a}
\rho=\rho_0(\frac{R_0}{R})^{2k+2},
\end{eqnarray}
respectively, where $R_0$ is the position of the shell at $t=t_0$
and $\rho_0$ is the density of matter on the shell at position
$R_0$. It is mentioned here that in case of finite $n$, energy
density diverges at
$R=R_0(\frac{k}{k+{\rho_0}^{\frac{-1}{n}}})^\frac{n}{2}$.

Twice squaring Eq.(\ref{15}), we obtain equation of motion of the
shell
\begin{equation}\label{19}
\dot{R}^2+V_{eff}(R)=0,
\end{equation}
where the effective potential $V_{eff}(R)$ is
\begin{equation}\label{20}
V_{eff}(R)=\frac{1}{2}(L_++L_{-})
-\frac{(L_+-L_{-})^2}{({\kappa}{\rho}R)^2}
-\frac{1}{16}({\kappa}{\rho} R)^2.
\end{equation}
\begin{figure}
\center\epsfig{file=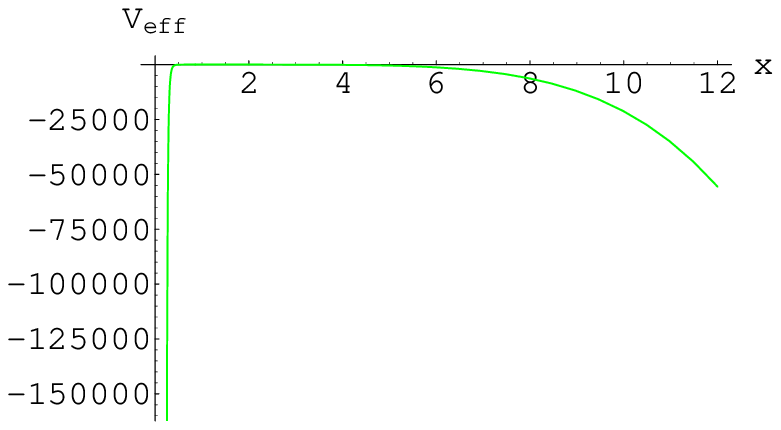, width=0.45\linewidth}
\epsfig{file=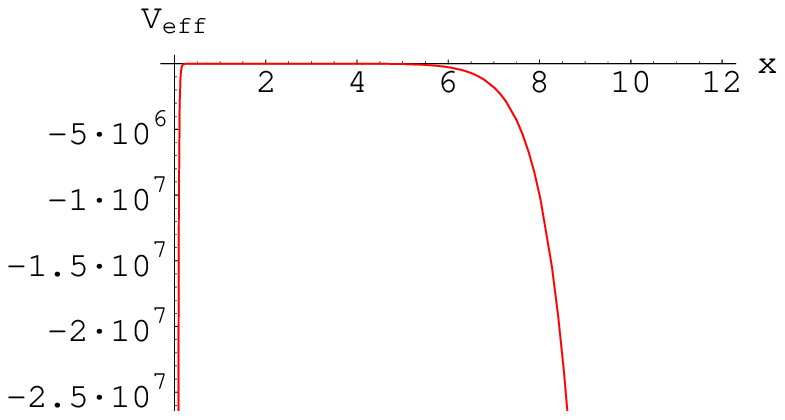, width=0.45\linewidth} \caption{(Color online)
Both graphs represent the effective potential for the polytropic
matter shell (\ref{22}). The left graph corresponds to $n=30$ and
$k=2$ while the right graph corresponds to $n=-30$ and $k=2.$ For
both graphs the values of the parameters are $M_-=0,
M_+=R_0=\rho_0=Q=1,\kappa=8\pi$. These values of the parameters will
remain the same for each graph while the extra parameters will be
mentioned.} \center\epsfig{file=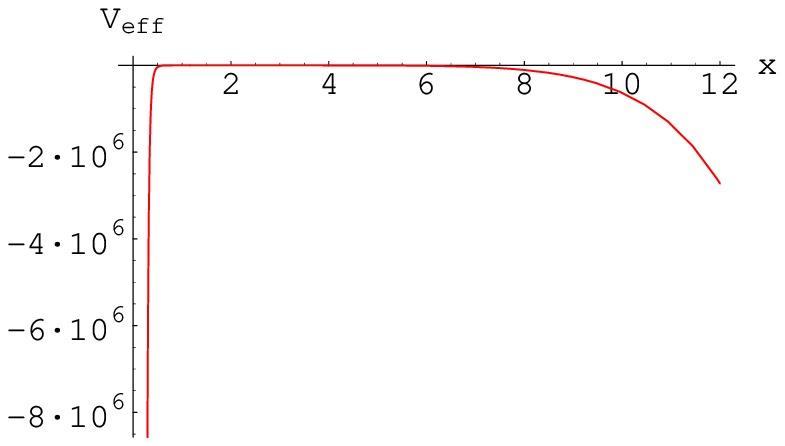, width=0.45\linewidth}
\epsfig{file=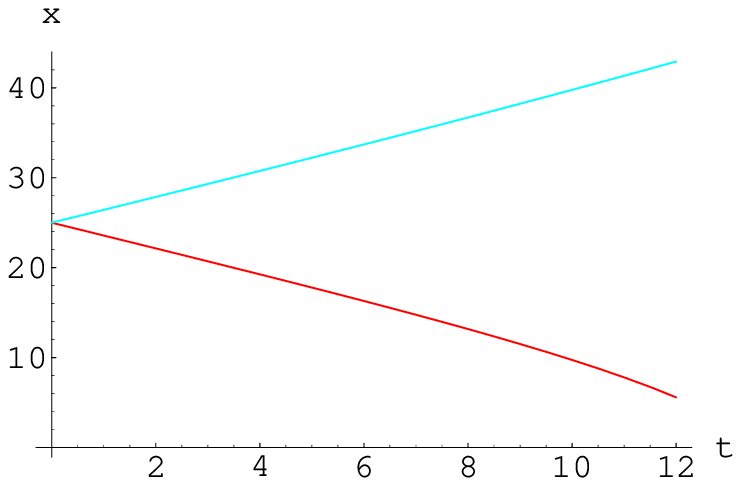, width=0.45\linewidth}\caption{(Color online)
The left graph shows the effective potential for the perfect fluid
shell (\ref{22a}) with $k=2$. The right graph represents the
increase and decrease in the shell radius with the increase of time
for ${x}(0)=25$.}
\end{figure}
Using $x=\frac{R}{R_0}$ and $\tau=\frac{t}{R_0}$, this equation
reduces to the following form
\begin{equation}\label{20a}
\dot{x}^2+V_{eff}(x)=0.
\end{equation}
The corresponding effective potential from Eqs.(\ref{18}) and
(\ref{20}) for finite $n$ is
\begin{equation}\label{22}
V_{eff}(x)=1-\frac{\varepsilon_{+}}{x}+\frac{\tilde{Q}^2}{x^2}
-\frac{{\varepsilon^2_-}{\eta}^2}{4x^4}(x^{\frac{2}{n}}-d)^{2n}
-\frac{x^2}{\eta^2{(x^{\frac{2}{n}}-d)^{2n}}},
\end{equation}
where $\varepsilon_\pm=\frac{(M_+\pm
M_-)}{R_0},~\tilde{Q}=\frac{Q}{R_0},~d=\frac{k}{k+\rho_0^\frac{-1}{n}},~
\eta=4\frac{(k+\rho_0^\frac{-1}{n})^n}{\kappa R_0}$. For infinite
$n$ (perfect fluid), it turns out to be
\begin{equation}\label{22a}
V_{eff}(x)=1-\frac{\varepsilon_{+}}{x}+\frac{\tilde{Q}^2}{x^2}-\frac{\bar{\eta}^2}{x^{2+4k}}
-\frac{\varepsilon^2_-{x^{4k}}}{4\bar{\eta}^2},
\end{equation}
where $\bar{\eta}=\frac{1}{4}{\kappa}\eta {R_0}^{(1-4k)}$.

Now we discuss Eqs.(\ref{20a})-(\ref{22a}) graphically following the
recent papers.$^{{27, ~28})}$  Fig. 1 describes the behavior of
effective potential (\ref{22}) for the collapsing polytropic matter
with finite $n$ and shell initial data. The left graph in Fig. 2
indicates the behavior of effective potential (\ref{22a}) for the
perfect fluid shell depending on the EoS parameter and shell initial
data. All these graphs show that $V_{eff}\leqslant0$, thus
Eq.(\ref{20a}) implies that motion is possible as
$\dot{x}^2\geqslant0$. The left graphs in both figures show that the
effective potential increases from $-\infty$ to $0$ and then
decreases from $0$ to finite negative value. This implies that
expanding or collapsing shell of matter comes to static state and
then expands or collapse. The right graph in Fig. 1 represents the
following three phases:
\begin{enumerate}
\item Initially $V_{eff}\rightarrow -\infty$ as
$x\rightarrow0$, in this case, polytropic matter shell will expand
to infinity for large initial radius or collapses to zero radius,
forming a black hole or naked singularity. In both cases bouncing
would occur if initial shell velocity is negative or positive,
respectively.
\item When $V_{eff}\rightarrow0$ for $x>0$. The
matter shell attains the non-static equilibrium state as $x$
increases.
\item $V_{eff}\rightarrow-\infty$ for some values of $x$.
This implies that static shell comes to the state of expansion or
collapse. In case of no crossing to $x$-axis, the shell collapses to
zero or expands to infinity depending on the choice of the shell
initial data.
\end{enumerate}

The effect of charge parameter on the dynamics of the shell can be
determined by the shell equation of motion given by
\begin{equation}\label{19}
\dot{x}=\pm\sqrt{-1+\frac{\varepsilon_{+}}{x}-\frac{\tilde{Q}^2}{x^2}
+\frac{{\varepsilon^2_-}{\eta}^2}{4x^4}(x^{\frac{2}{n}}-d)^{2n}
+\frac{x^2}{\eta^2{(x^{\frac{2}{n}}-d)^{2n}}}}.
\end{equation}
Here $\pm$ correspond to expansion and collapse of the shell,
respectively. It follows from this equation that the presence of the
charge term (Coulomb repulsive force) decreases the velocity of the
expanding or collapsing shell. Also, we would like to mention here
that this behavior of charge parameter is the same for the cases of
finite and infinite $n$. In this case, the shell motion is slow as
compared to the Schwarzschild spacetime.

Since the equation of motion (\ref{20a}) is nonlinear, its exact
solution is impossible but can be solved numerically. The numerical
solution of this equation for suitable choice of initial data and
for some initial value of shell radius gives the behavior of the
shell radius with respect to time shown in right Fig. 2. This shows
that radius decreases with the increase of time which is the strong
argument for a shell to collapse. Also, radius is an increasing
function of time which shows expansion.

\section{Non-commutative Correction to Thin Shell Collapse}

In this section, we study gravitational collapse in the
non-commutative RN geometry. First, we discuss the effects of the
non-commutative factor $\Theta$ on junction conditions which provide
the equations of motion. The spacetime non-commutativity can be
encoded by the following relation$^{{21})}$
\begin{equation}\label{23a}
[x^\mu,x^\nu]=\iota{\Theta}^{\mu\nu},
\end{equation}
where ${\Theta}^{\mu\nu}$ is an anti-symmetric matrix that
determines the spacetime cell discretization as $\hbar$ (Planck's
constant) discretizes the phase space. There are different
approaches to the non-commutative field theory out of which one is
based on the $\star$-product and another on the coordinate coherent
state formalism. Recently, following the second approach, Smailagic
and Spallucci$^{{29})}$ have shown that the problems of Lorentz
invariance and unitary (arising in the $\star$-product approach) can
be solved by considering ${\Theta}^{\mu\nu}={\Theta}
diag(\epsilon_{ij},\epsilon_{ij}......)$, where ${\Theta}$ is
constant with dimensions of length squared. Also, the coordinate
coherent state modifies the Feynman propagators. Thus it is believed
that non-commutativity removes the singularities (divergences)
appearing in GR. In GR, the metric field is a geometrical structure
and curvature (presence of matter) measures its strength. Since
non-commutativity is the fundamental property of the metric, so it
affects gravity via curvature. This implies that non-commutativity
influences the matter energy-momentum tensor. Thus the geometric
part of the field equations is left unchanged and modification is
made in the energy-momentum tensor. Following this philosophy, we
shall make only modification in matter part of the junction
conditions and leave the geometric part unchanged.

The line element for this metric is$^{{23})}$
\begin{equation}\label{23}
(ds)_\pm^2=L_\pm dT^2-\frac{1}{L}_\pm
dR^2-R^2(d\theta^2+\sin^2{\theta}d\phi^2),
\end{equation}
where
\begin{eqnarray*}
L_\pm(R)&=&1-\frac{4M_\pm}{{R}
\sqrt{\pi}}\gamma\left(\frac{3}{2};\frac{R^2}{4\Theta}
\right)+\frac{Q^2}{\pi R^2}\gamma^2\left(\frac{1}{2};
\frac{R^2}{4\Theta}\right)-\frac{Q^2}{\pi
R\sqrt{2\Theta}}\gamma\left(\frac{1}{2};\frac{R^2}{2\Theta}
\right)\\
&+&\frac{Q^2}{{\pi}R}\sqrt{\frac{2}{\Theta}}\gamma(\frac{3}{2};\frac{R^2}{4\Theta})
\end{eqnarray*}
and lower incomplete gamma function is defined by
\begin{equation}\label{a}
\gamma\left(\frac{a}{b};x \right)=\int^x_0
\frac{dt}{t}t^{\frac{a}{b}} e^{-t}.
\end{equation}
In the commutative limit
$\frac{R}{\sqrt{\Theta}}\longrightarrow\infty$, i.e.,
${\Theta}\rightarrow0$, Eq.(\ref{23}) reduces to conventional RN
metric (\ref{1}) (for detail, see the properties of gamma function
in the appendix $^{{22})}$.

Here we assume the smeared gravitating source in the non-commutative
geometry and use the modified energy density and pressure$^{{26})}$
$\rho_m=\rho_s+\rho_\Theta$ and $p_m=p_s+p_\bot$ respectively. The
quantities $\rho_s$ and $p_s$ are the energy density and pressure of
the shell used in the previous section while $\rho_\Theta$ and
$p_\bot$ are the energy density and pressure of the smeared source
in the non-commutative theory. Since pressure and density of matter
are effected by the non-commutative factor $\Theta$, so equation of
state is also effected by this factor. Recently, Bertolami and Zarro $^{{25})}$ have pointed
out this effect by studying the astrophysical
objects in the context of non-commutativity. For non-commutative
energy density, $\rho_\Theta$ and pressure, $p_\bot$, Eq.(\ref{16})
can be written as
\begin{equation}\label{25}
\frac{d\rho_\Theta}{d log R}+2(\rho_\Theta+p_\bot)=0.
\end{equation}
It is well-known that $\rho_\theta$ and $p_\bot$ satisfy the
following relation$^{{26})}$
\begin{equation}\label{26}
p_\bot=-(1-\frac{R^2}{4\Theta})\rho_\Theta.
\end{equation}
Thus the non-commutative energy density is
$\rho_\Theta=\bar{\rho}e^{{-(\frac{R^2-R_0^2}{4\Theta})}}$, where
$\bar{\rho}$ is the value of $\rho_\Theta$ at position $R_0$. It is
interesting to note that when $R \rightarrow0$ or $\Theta
\rightarrow\infty$, this matter acts as a matter of constant
density. Thus the modified energy density for finite $n$ is
\begin{equation}\label{27}
\rho_m=\{(k+{\rho_0}^{\frac{-1}{n}})(\frac{R}{R_0})^{\frac{2}{n}}-k\}^{-n}+
\bar{\rho}e^{{-(\frac{R^2-R_0^2}{4\Theta})}}.
\end{equation}

Also, for perfect fluid, we have
\begin{equation}\label{28}
\rho_m=\rho_0(\frac{R_0}{R})^{2k+2}+\bar{\rho}~
e^{{-(\frac{R^2-R_0^2}{4\Theta})}}.
\end{equation}
Equation of motion of the shell is
\begin{equation}\label{29}
\dot{R}^2+V_{eff}(R)=0,
\end{equation}
where
\begin{equation}\label{30}
V_{eff}(R)=\frac{1}{2}(L_++L_{-})-\frac{(L_+-L_{-})^2}{({\kappa}{\rho}_m
R)^2} -\frac{1}{16}({\kappa}{\rho}_m R)^2.
\end{equation}
Using $x=\frac{R}{R_0},~\tau=\frac{t}{R_0}$ as in the previous case,
we get the modified energy density and effective potential for
finite $n$ as follows
\begin{eqnarray}\label{30a}
\rho_m&=&\{(k+{\rho_0}^{\frac{-1}{n}})x^{\frac{2}{n}}-k\}^{-n}+
\bar{\rho}e^{{-R_0^2(\frac{x^2-1}{4\Theta})}},\\\label{31}
V_{eff}(x)&=&1-\frac{\varepsilon_{+}}{x}+\frac{\tilde{Q}^2}{x^2}-\frac{4\varepsilon^2_-}{\kappa^2R_0^2x^4}
\left(\{(k+{\rho_0}^{\frac{-1}{n}})x^{\frac{2}{n}}-k\}^{-n}+
\bar{\rho}e^{{-R_0^2(\frac{x^2- 1}{4\Theta}})}\right)^{-2}
\nonumber\\
&-&\frac{R_0^2\kappa^2x^2}{16}\left(\{(k+{\rho_0}^{\frac{-1}{n}})x^{\frac{2}{n}}-k\}^{-n}+
\bar{\rho}e^{{-R_0^2(\frac{x^2- 1}{4\Theta}})}\right)^{2}.
\end{eqnarray}
Also, using Eqs.(\ref{28}) and (\ref{30}), the effective potential
for the perfect fluid is
\begin{eqnarray}\label{31a}
V_{eff}(x)&=&1-\frac{\varepsilon_{+}}{x}+\frac{\tilde{Q}^2}{x^2}-\frac{4\varepsilon^2_-}{\kappa^2R_0^2x^4}
\{\rho_0(\frac{1}{x})^{(2k+2)}+\bar{\rho}e^{-R_0^2(\frac{x^2-1}{4\Theta})}
\}^{-2}\nonumber\\
&-&\frac{R_0^2\kappa^2x^2}{16}
\{\rho_0(\frac{1}{x})^{(2k+2)}+\bar{\rho}e^{-R_0^2(\frac{x^2-1}{4\Theta})}
\}^{2}.
\end{eqnarray}
\begin{figure}
\center\epsfig{file=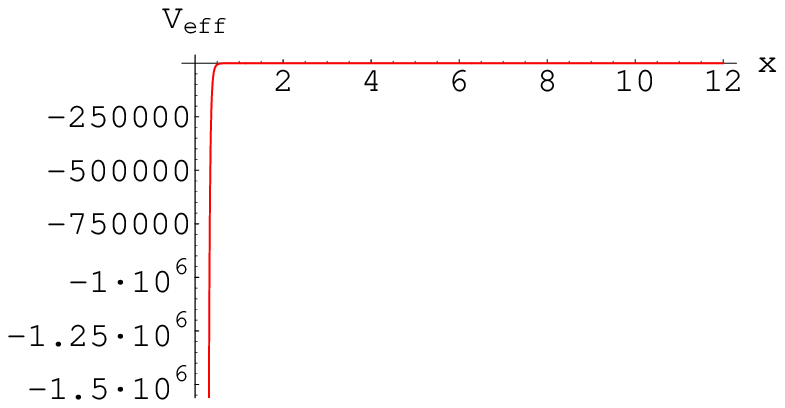,
width=0.45\linewidth}\epsfig{file=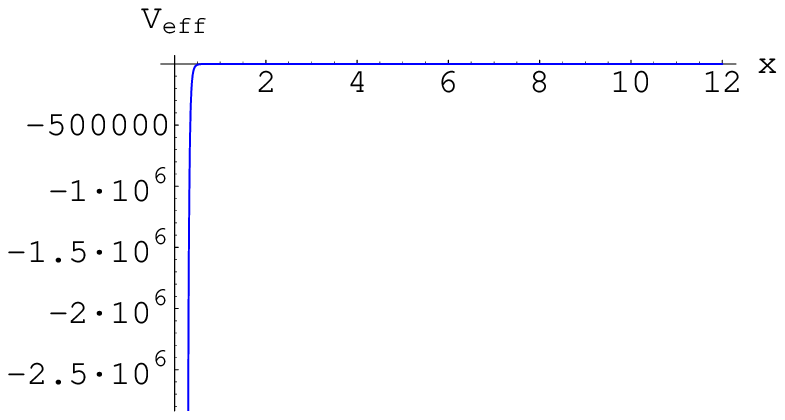, width=0.45\linewidth}
\caption{(Color online) Both graphs show the effective potential for
the polytropic matter shell in non-commutative background
(\ref{31}). The left graph corresponds to $n=30$ and $\Theta=4$
while the right graph to $n=30$ and $\Theta=8$. For both graphs
$k=2, \bar{\rho} =1$.}
\end{figure}
\begin{figure}
\center\epsfig{file=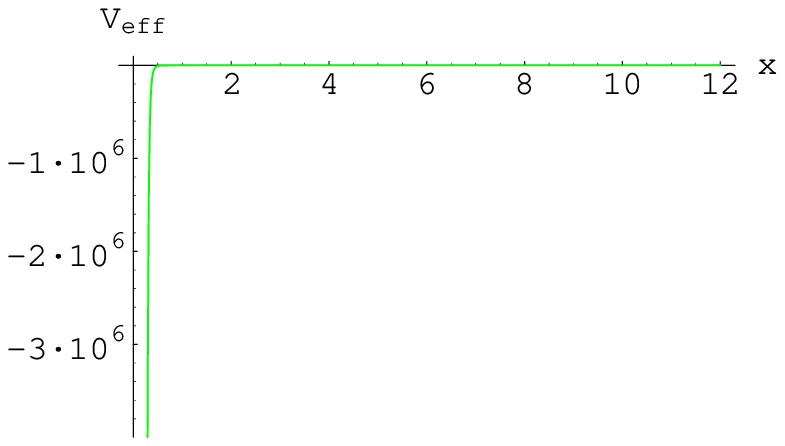,
width=0.45\linewidth}\epsfig{file=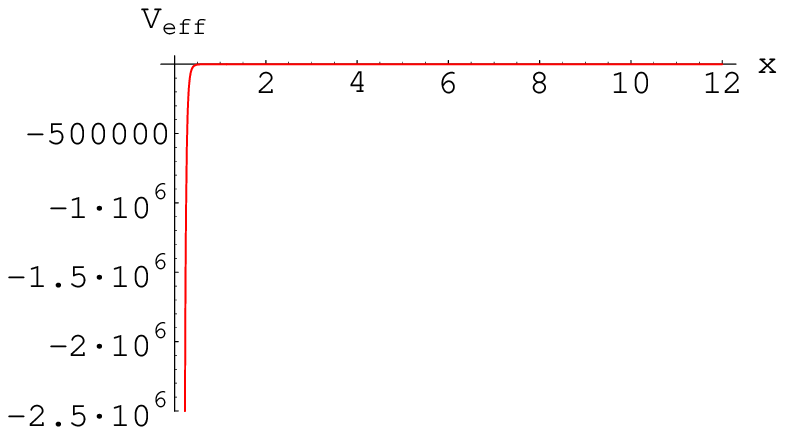, width=0.45\linewidth}
\caption{(Color online) These indicate the effective potential for
the polytropic matter shell in non-commutative background
(\ref{31}). The left graph corresponds to $n=30$ and $\Theta=12$
while the right graph to $n=-30$ and $\Theta=4$.}
\end{figure}
\begin{figure}
\center\epsfig{file=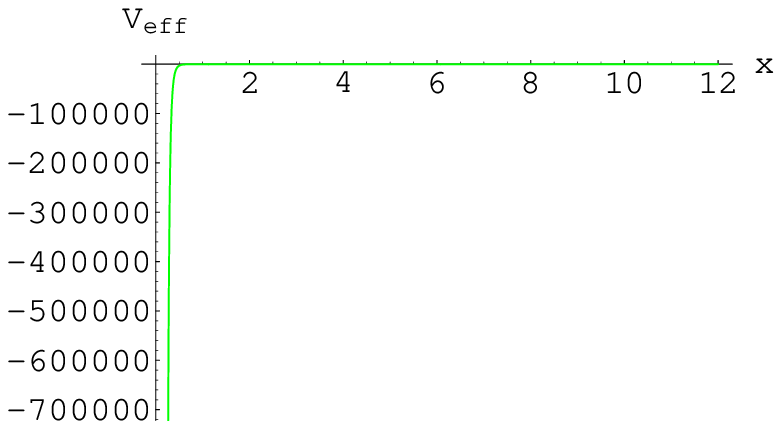,
width=0.45\linewidth}\epsfig{file=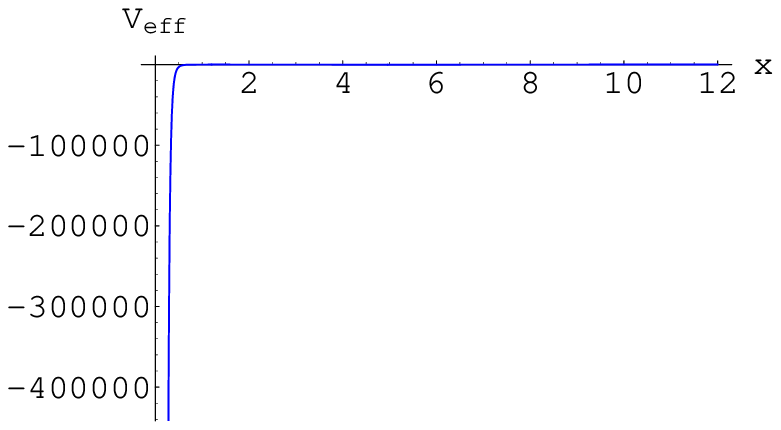, width=0.45\linewidth}
\caption{(Color online) The left and right graphs represent the
effective potential (\ref{31}) corresponding to $n=-30,~\Theta=8$
and $n=-30$,~ $\Theta=12$.}
\end{figure}
\begin{figure}
\center \epsfig{file=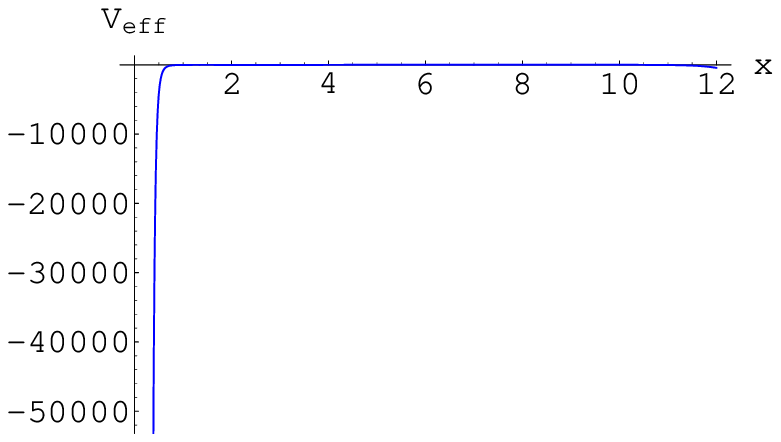,
width=0.45\linewidth}\epsfig{file=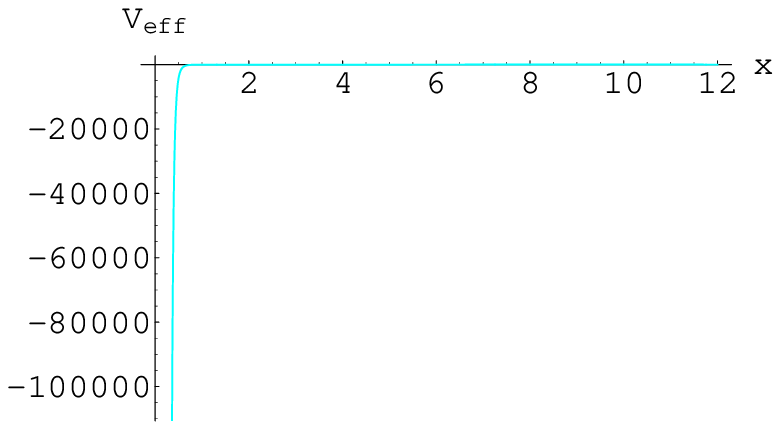, width=0.45\linewidth}
\caption{(Color online) The effective potential (\ref{31})
corresponding to $\Theta=4$ and $\Theta=8$ respectively.}
\end{figure}
\begin{figure}\center \epsfig{file=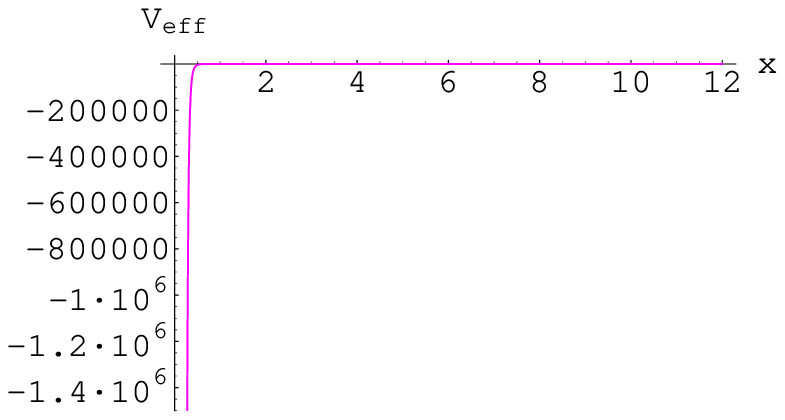, width=0.45\linewidth}
\epsfig{file=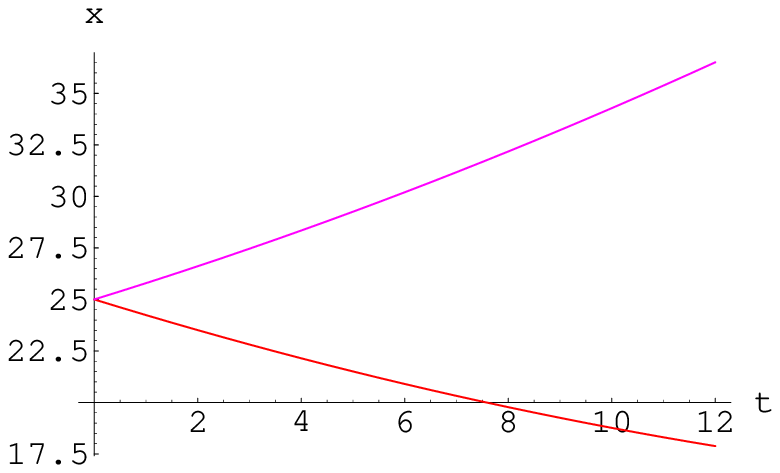, width=0.45\linewidth} \caption{(Color online)
The effective potential of the perfect fluid shell corresponding to
$\Theta=12$ is shown in the left graph. While the right graph shows
that for expanding and collapsing shell radius is increasing and
decreasing function of $t$, respectively.}
\end{figure}

Now we discuss the behavior of Eqs.(\ref{29})-(\ref{31a})
graphically. It follows from Figs. 4-6 that the effective potential
(36) increases from negative to zero for polytropic matter with
varying ${\Theta}$ and fixed $k$. The similar behavior of the
effective potential (37) for the perfect fluid shell is shown in the
left Fig. 7. It is mentioned here that we have only considered the
case for which $k>0$ in order to exclude the possibility of the
exotic matter (dark energy) shell for which $k<0$. This implies that
depending on the choice of initial data, the shell continuously
expands or collapses to a finite size then comes to rest position.
After a shell attains a last stage of rest position, it has no
capability to re-expand or re-collapse to zero size. All the graphs
in non-commutative background have $V_{eff}\rightarrow-\infty$ at
$x>0$, neither of them is divergent at $x=0$. This confirms that
non-commutativity can measure the short distances up to the order of
Plank's length scale. This can be seen by investigating the horizon
radius and point of singularity where density diverges. The exact
solution of nonlinear equation (\ref{29}) is impossible. As before,
we solve this equation by using numerical technique with initial
condition. The behavior of the shell radius in this case is shown in
right graph of Fig. 7.

The black hole horizon can be found by solving
\begin{eqnarray}\label{33}
&1&-\frac{4M_\pm}{{xR_0 } \sqrt{\pi}}\gamma\left(\frac{3}{2};
\frac{(xR_0)^2}{4\Theta} \right)+\frac{Q^2}{\pi
(xR_0)^2}\gamma^2\left(\frac{1}{2}; \frac{(xR_0)^2}{4\Theta}
\right)-\frac{Q^2}{\pi (xR_0)\sqrt{2\Theta}}
\nonumber\\
&\times& \gamma\left(\frac{1}{2}; \frac{(xR_0)^2}{2\Theta}
\right)+\frac{Q^2}{{\pi}(xR_0)}\sqrt{\frac{2}{\Theta}}
\gamma\left(\frac{3}{2};\frac{(xR_0)^2}{4\Theta}\right)=0.
\end{eqnarray}
For $R_0=1,~Q=1,~M_+=1,~M_-=0$ and taking initially $x_h=0.1$, the
position of the horizon by iterative method is
\begin{eqnarray}\label{33}
x_h&=&1.35862;\quad \quad  \Theta=4,\nonumber\\
x_h&=&1.70161;\quad \quad \Theta=8,\nonumber\\
x_h&=&1.94338;\quad \quad \Theta=12.
\end{eqnarray}
For finite $n$, the modified energy density (\ref{30a}) as well as
the effective potential (\ref{31}) for polytropic matter shell in
non-commutative case are singular at
$x_s=\frac{k^{\frac{n}{2}}}{(k+{\rho_0^{\frac{-1}{n}}})^{\frac{n}{2}}}$.
Although this is independent of $\Theta$ but it is the only value of
$x$ at which modified energy density and effective potential
diverge. Further, all graphs of the effective potential for
polytropic matter in Fig. 3-6 for non-commutative case imply that
$V_{eff}$ diverges negatively at $x>0$, while the right graph of
Fig. 1 in commutative case for polytropic matter imply that
$V_{eff}\rightarrow -\infty$ at $x=0$. This means that the
non-commutative parameter $\Theta$ has shifted the singularity from
$x=0$ to $x>0$. Hence, for the values of the parameters, $k,~n$ and
$\rho_0$ used for the solutions previously, we get $x_s=0.00228365$
at which polytropic matter shell in non-commutative case becomes
singular. From the values of $x_h$ and $x_s$, we conclude that
"shell radii are greater than the singular point (where density
diverges)" i.e., horizon covers the singularity at point
$x_s=0.00228365$ which leads to the formation of black hole as the
final fate of the collapse as shown in Fig. 8.

For the case of infinite $n$, the energy density as well as the
effective potential diverge at $x=0$. Thus for each value of
$\Theta$, the corresponding values of horizon radius are greater
than zero, hence a singular shell of zero radius seems to be covered
by the horizon radius. Consequently, we can say that perfect fluid
shell collapse in non-commutative geometry always ends as a black
hole. Hence in non-commutativity, polytropic matter shell collapses
to a circle of small non-zero radius, while perfect fluid shell
collapses to zero radius. The clear effects of $\Theta$ appear in
the presence of generic polytropic matter.
\begin{figure}
\center \epsfig{file=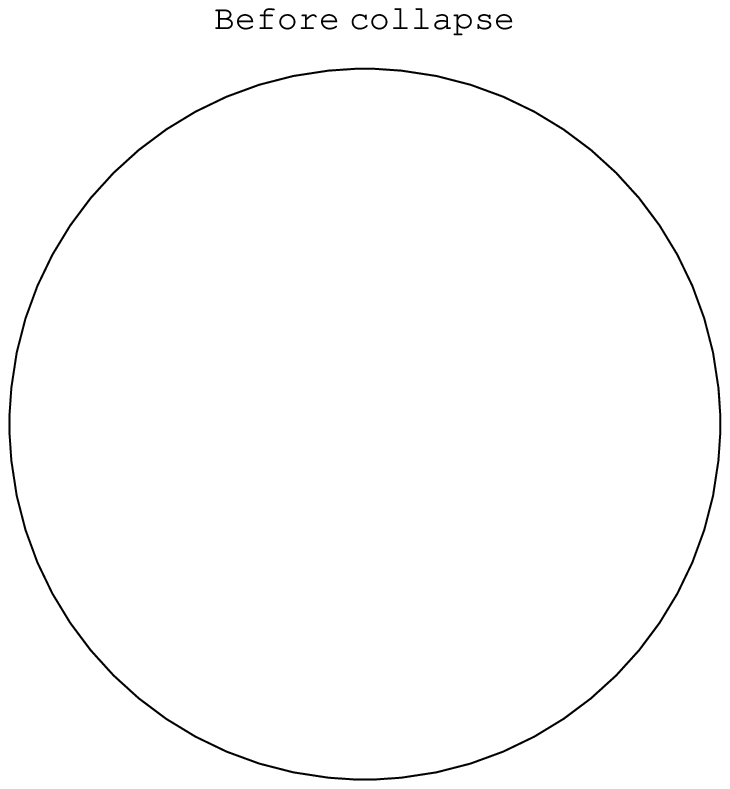,
width=0.45\linewidth}\epsfig{file=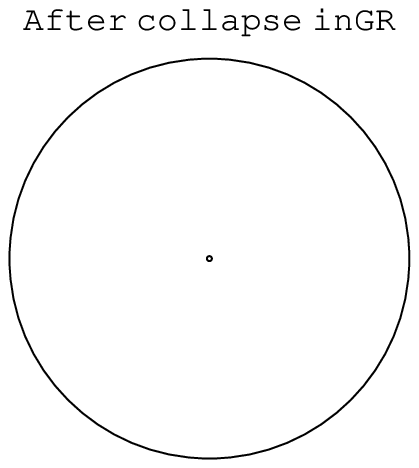,
width=0.3\linewidth}\epsfig{file=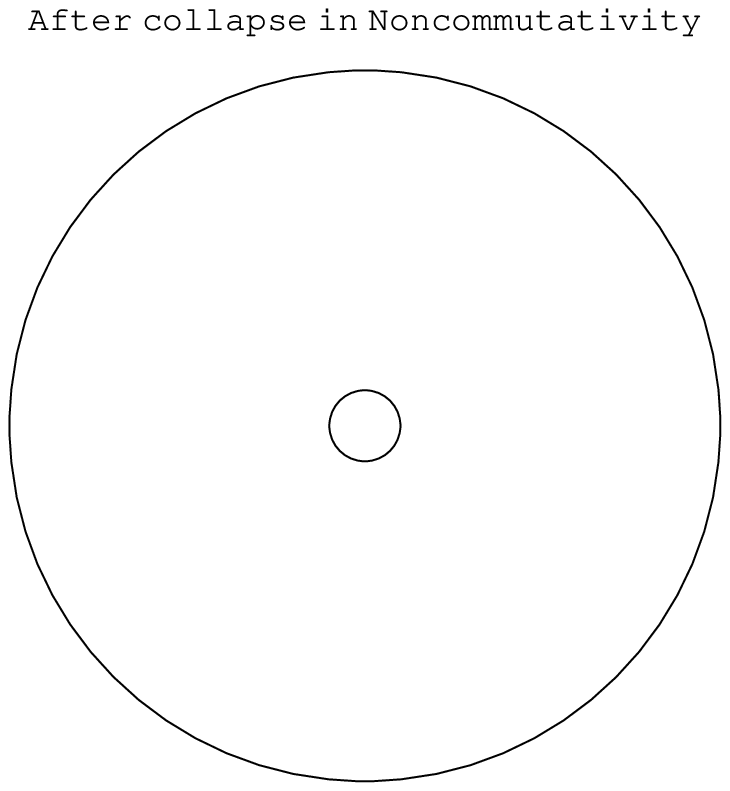, width=0.3\linewidth}
\caption{Note that shell before collapse is composed of matter,
after collapse in GR matter is contracted to a point, while in
non-commutative approach matter is contracted within an inner small
circle. According to GR, the shell collapses to zero radius leaving
behind event horizon while in non-commutativity, it collapses to
non-zero radius interior to the event horizon.}
\end{figure}

\section{Concluding Remarks}

This paper is devoted to study the gravitational collapse of
polytropic matter shell collapse in non-commutative RN geometry.
Using the Israel junction conditions between the interior and
exterior regions of a star, we have formulated equation of motion of
the shell. It has been found that solution of this equation
represents collapse and expansion depending on the choice of initial
data. Further, by taking the polytropic index $n$ finite and
infinite (perfect fluid), the effective potential of the system has
been discussed in detail. The behavior of the effective potential
for the polytropic matter with finite $n$ shows that the matter can
undergo an expansion, collapse or bouncing depending on the choice
of initial data. Similarly, the effective potential of the system
for infinite $n$ (perfect fluid) predicts different stages of
dynamics of the shell depending on the choice of initial data.

In order to understand the effects of non-commutative factor
$\Theta$ on the gravitational collapse, a smeared source in the
non-commutative RN spacetime has been taken. We have adopted the
non-commutative approach$^{{26})}$ to see the effects of the
non-commutative factor $\Theta$ on the collapse. Using the modified
matter components, the shell equations of motion for polytropic
matter with finite and infinite $n$ have been derived. In this
approach, we have found that in the presence of $\Theta$ an
initially an expanding or collapsing matter shell comes to static
equilibrium. It cannot further, re-expand to infinity or re-collapse
to zero size shown in Figs. 3-7. In other words in non-commutative
the end state is a BH with some non-zero horizon radius.

It has been found that the charge parameter $Q$ effects the velocity
of the expanding or collapsing shell but does not play a dominant
role to bring a shell to static equilibrium. For charged to be
dominant it is necessary that effective potential must vanish in
general, so that we get velocity $\dot{x}=0$. Since it is not
possible in general, so charge only effects the dynamics of the
shell but doesnot prevent the collapse. Further, the solution of
$g_{00}=0$ for the non-commutative RN spacetime gives the horizon
radius. For the larger value of $\Theta$ factor, it has been found
that the horizon radius covers the singularity where density is
undefined for the case of polytropic matter with finite $n$. For any
non-zero value of $\Theta$, perfect fluid always collapse to form a
black hole. Thus in the non-commutative approach, there is a
validity of the CCH depending on the choice of initial data.

After the work of Oh and Park$^{{26})}$, the present work is the
second step towards the thin shell collapse in non-commutative
theory. Although both papers involve the numerical solutions
however, it is expected that this kind of study would lead to find
an exactly solvable collapsing model in non-commutative theory of
gravity as Oppenheimer-Synder$^{{11})}$ classical collapsing model
in GR. The thin shell formalism was used for a wide class of
cosmological and and astrophysical problems $^{{30},{31})}$ which
could be extended by using non-commutative approach.

\vspace{0.25cm}

{\bf Acknowledgments}

\vspace{0.25cm}

We would like to thank the Higher Education Commission, Islamabad,
Pakistan for its financial support through the {\it Indigenous Ph.D.
5000 Fellowship Program Batch-IV}. Also, we highly appreciate the
fruitful comments of the anonymous referee.
\\\\
1) R. Penrose: Riv. Nuovo Cimento \textbf{1} (1969) 252.\\\
2) S. W. Hawking and  G. F. R. Ellis: \emph{The Large Scale
Structure of Spacetime}
(Cambridge University Press, Camgridge, U.K., 1975).\\\
3) K. S. Virbhadra, D. Narasimha and M. S. Chitre: Astron.
Astrophys.
\textbf{337} (1998) 1.\\\
4) K. S. Virbhadra and  G. F. R. Ellis: Phys. Rev.
\textbf{D 62} (2000) 084003.\\\
5) K. S. Virbhadra and G. F. R. Ellis: Phys. Rev.
\textbf{D 65} (2002) 103004.\\\
6) C. M. Claudel, K. S. Virbhadra and  G. F. R. Ellis: J. Math.
Phys. \textbf{42} (2001) 818.\\\
7) K. S. Virbhadra and C. R. Keeton: Phys. Rev.
\textbf{D 77} (2008) 124014.\\\
8) K. S. Virbhadra: Phys. Rev. \textbf{D 60} (1999) 104041.\\\
9) K. S. Virbhadra: Phys. Rev. \textbf{D 79} (2009) 083004.\\\
10) W. Israel: Nuovo Cimento \textbf{B 44} (1966) 1.\\\
11) J. R. Oppenheimer and H. Snyder: Phys. Rev. \textbf{56} (1939) 455.\\\
12) P. R. C. T. Pereira and A. Wang: Phys. Rev. {\bf D 62} (2000) 124001.\\\
13) M. Sharif and Z. Ahmad: Int. J. Mod. Phys. \textbf{A 23} (2008) 181.\\\
14) M. Sharif and K. Iqbal: Mod. Phys. Lett.
\textbf{A 24} (2009) 1533.\\\
15) M. Sharif and G. Abbas: Gen. Relativ. Gravity \textbf{43} (2011) 1179.\\\
16) V. de La Cruz and W. Israel: Nuovo Cimento \textbf{A 51} (1967) 744.\\\
17) K. Kuchar: Czechoslovak J. Physics Section \textbf{B 18} (1968) 435.\\\
18) J. E. Chase: Nuovo Cimento \textbf{B 67} (1970) 136.\\\
19) D. G. Boulware: Phys. Rev. \textbf{D 8} (1973) 2363.\\\
20) S. Doplicer, K. Fredenhagen and J. E. Robert: Commun. Math. Phys. \textbf{172} (1995) 187.\\\
21) P. Nicolini, A. Smailagic, and E. Spallucci: Phys. Lett. \textbf{B 632} (2006) 547.\\\\\\
22) L. Modesto and P. Nicolini: Phys. Rev. \textbf{D 82} (2010) 104035.\\\
23) C. Bastos, O. Bertolami, N. C. Dias and J. N. Prata: Phys. Rev. \textbf{D 80} (2009) 124038.\\\
24) C. Bastos, O. Bertolami, N. C. Dias and J. N. Prata: Phys. Rev. \textbf{D 84} (2011) 024005.\\\
25) O. Bertolami and C. D. A. Zaro: Phys. Rev. \textbf{D 81} (2010) 025005.\\\
26) J. J. Oh and C. Park: JHEP \textbf{1003} (2010) 86.\\\
27) R. B. Mann and J. J. Oh: Phys. Rev. \textbf{D 74} (2006) 124016.\\\
28) R. B. Mann, J. J. Oh and P. Mu-In: Phys. Rev. \textbf{D 79} (2009) 064005.\\\
29) A. Smailagic and E. Spallucci: J. Phys. \textbf{A 36} (2003) L 467.\\\
30) C. Barrabes and W. Israel: Phys. Rev. \textbf{D 43} (1991) 1129.\\\
31) A. V. Berezin, V. A. Kuzmin and I. I. Tkachev: Phys. Lett.
\textbf{B 120} (1983) 91.
\end{document}